\documentclass[journal]{IEEEtran}

\usepackage{xcolor,soul,framed} 

\colorlet{shadecolor}{yellow}
\usepackage[pdftex]{graphicx}
\graphicspath{{../pdf/}{../jpeg/}}
\DeclareGraphicsExtensions{.pdf,.jpeg,.png}

\usepackage[cmex10]{amsmath}
\usepackage{array}
\usepackage{mdwmath}
\usepackage{mdwtab}
\usepackage{eqparbox}
\usepackage{url}
\usepackage{cite}
\usepackage{amsmath,amssymb,amsfonts}
\usepackage{algorithm}
\usepackage{algpseudocode}
\usepackage{graphicx}
\usepackage{textcomp}
\usepackage{tcolorbox}
\usepackage{xcolor}
\usepackage{listings}
\usepackage{url}
\usepackage{array}
\usepackage{tablefootnote}
\usepackage{bbding}
\usepackage{makecell} 
\usepackage{subcaption}
\usepackage{booktabs}
\usepackage{multirow}
\usepackage{balance}
\usepackage{soul}
\usepackage[edges]{forest}
\usepackage{tikz}
\usetikzlibrary{arrows.meta}
\usetikzlibrary{automata, positioning}

\soulregister\cite7          
\soulregister\ref7
\soulregister\pageref7

\definecolor{codegreen}{rgb}{0,0.6,0}
\definecolor{backcolour}{rgb}{0.95,0.95,0.92}
\definecolor{codegray}{rgb}{0.5,0.5,0.5}
\definecolor{codepurple}{rgb}{0.58,0,0.82}
\lstdefinestyle{listingstyle}{
    backgroundcolor=\color{white},   
    commentstyle=\color{codegreen},
    keywordstyle=\color{magenta},
    numberstyle=\tiny\color{codegray},
    stringstyle=\color{codepurple},
    basicstyle=\ttfamily\footnotesize,
    breakatwhitespace=false,         
    breaklines=true,                 
    captionpos=b,                    
    keepspaces=true,                 
    numbers=left,                    
    numbersep=5pt,                  
    showspaces=false,                
    showstringspaces=false,
    showtabs=false,                  
    tabsize=1,
    frame=single
}
\def\BibTeX{{\rm B\kern-.05em{\sc i\kern-.025em b}\kern-.08em
    T\kern-.1667em\lower.7ex\hbox{E}\kern-.125emX}}

\begin{document}
\bstctlcite{IEEEexample:BSTcontrol}
    \title{InsightQL: Advancing Human-Assisted Fuzzing with a Unified Code Database and Parameterized Query Interface}
  
    \author{Wentao~Gao,
            Renata~Borovica\mbox{-}Gajic,
            Sang~Kil~Cha,
            Tian~Qiu,
            and~Van\mbox{-}Thuan~Pham
    \thanks{W.~Gao is with The University of Melbourne, Australia (e-mail: wentaog1@student.unimelb.edu.au).}%
    \thanks{R.~Borovica\mbox{-}Gajic is with The University of Melbourne, Australia (e-mail: renata.borovica@unimelb.edu.au).}%
    \thanks{S.~K.~Cha is with The Korea Advanced Institute of Science and Technology
, Korea (e-mail: sangkilc@kaist.ac.kr).}%
    \thanks{T.~Qiu is with The University of Melbourne, Australia (e-mail: tq@student.unimelb.edu.au).}%
    \thanks{V.\mbox{-}T.~Pham is with The University of Melbourne, Australia (e-mail: thuan.pham@unimelb.edu.au).}%
    }



\newcommand{\codeql}[1]{\textsc{CodeQL}}
\newcommand{\insightql}[1]{\textsc{InsightQL}}
\newcommand{\fuzzintrospector}[1]{\textsc{FuzzInstropector}}
\newcommand{\ossfuzz}[1]{\textsc{OSSFuzz}}

\newcommand{\TODO}[2]{\colorbox{yellow}{\textbf{TODO:}} \color{blue} \emph{#1} (\textbf{#2}) \color{black}}

\newcommand{\CHANGE}[1]{\colorbox{yellow}{\textbf{CHANGE:}}~\color{blue} #1 \color{black}}

\newcommand{\FIXME}[2]{\colorbox{red}{\textbf{FIXME:}} \color{red} \emph{#1} (\textbf{#2}) \color{black}}

\newcommand{\textbox}[1]{
    \begin{tcolorbox}[colback=gray!5!white,colframe=gray!100!black]
        #1
    \end{tcolorbox}
}

\maketitle

\begin{abstract}
Fuzzing is a highly effective automated testing method for uncovering software vulnerabilities. Despite advances in fuzzing techniques, such as coverage-guided greybox fuzzing, many fuzzers struggle with coverage plateaus caused by fuzz blockers, limiting their ability to find deeper vulnerabilities. Human expertise can address these challenges, but analyzing fuzzing results to guide this support remains labor-intensive. To tackle this, we introduce \insightql{}, the first human-assisting framework for fuzz blocker analysis. Powered by a unified database and an intuitive parameterized query interface, \insightql{} aids developers in systematically extracting insights and efficiently unblocking fuzz blockers. Our experiments on 14 popular real-world libraries from the FuzzBench benchmark demonstrate the effectiveness of \insightql{}, leading to the unblocking of many fuzz blockers and considerable improvements in code coverage (up to 13.90\%).
\end{abstract}


%
\IEEEpeerreviewmaketitle


\section{Introduction} \label{sec:intro}

Fuzzing is one of the most effective and efficient automated testing approaches for discovering software vulnerabilities \cite{miller1990empirical, manes2019art, bohme2020fuzzing, zhu2022fuzzing}. The technique comes in several forms (e.g., black-box, grey-box, white-box, dumb/smart fuzzing) and can generate millions of semi-valid random inputs within a few hours to detect anomalies in the System Under Test (SUT), far exceeding the scalability of manual testing.

Since its inception in 1990 \cite{miller1990empirical}, fuzzing has seen significant advancements in effectiveness, efficiency, and applicability—enabling faster bug discovery in a wide range of applications \cite{bohme2016coverage, bohme2017directed, pham2019smart, lyu2019mopt, lemieux2018fairfuzz, aschermann2019redqueen, fioraldi2020afl++, li2017steelix, peng2018t, wang2017skyfire, aschermann2019nautilus, pham2020aflnet, pan2024edefuzz, choi:oakland:2021, kim2023dafl, manes2020ankou}. Major technology companies such as Google, Microsoft, and Meta have developed and deployed fuzzing frameworks \cite{clusterfuzz, onefuzz}, leading to the discovery of thousands of vulnerabilities in widely-used systems including web browsers, operating systems, and open-source libraries.

However, our recent research \cite{gao2023beyond} indicates that fuzzing--particularly coverage-guided grey-box fuzzing \cite{bohme2020boosting, fioraldi2020afl++}--is reaching its limits. Fuzzers often hit a coverage plateau, even when many unexplored paths remain in the SUT, thereby failing to uncover potentially critical vulnerabilities. These limitations are often due to \emph{fuzz blockers}, which can be input-dependent or input-independent~\cite{gao2023beyond, fuzzintrospector}. For example, a fuzzer may struggle to generate an input that satisfies a specific condition in the SUT, such as a complex branch constraint or a required function argument.

Ideally, fuzzing should be enhanced to \emph{automatically} overcome such fuzz blockers and solve complex constraints—potentially with help from recent advances in Artificial Intelligence (AI) and Large Language Models (LLMs). However, this remains challenging due to AI’s limited understanding of SUT semantics and the scalability issues of symbolic execution, making \emph{human-assisted fuzzing a viable solution} \cite{shoshitaishvili2017rise, bohme2020fuzzing}.

Humans, such as software developers, testers, and security researchers, can assist fuzzers at different stages. In the typical three-phase fuzzing workflow (Preparation–Fuzzing–Result Analysis), humans are usually involved only in the first and last phases. This is referred to as \emph{Human-out-of-the-Loop (HOOTL)} fuzzing. A more advanced setup, known as \emph{Human-in-the-Loop (HITL)} fuzzing, has been increasingly promoted by both academia and industry \cite{bohme2020fuzzing}. In HITL fuzzing, humans are allowed and expected to assist the fuzzer during its main loop—especially when it encounters challenges like coverage plateaus \cite{shoshitaishvili2017rise}.

Several studies have demonstrated the usefulness of human support in both HITL and HOOTL setups \cite{shoshitaishvili2017rise, pham2019smart, aschermann2019nautilus, wang2019superion, aschermann2020ijon}. For instance, in HITL fuzzing, humans can directly provide additional seed inputs \cite{shoshitaishvili2017rise}. In HOOTL fuzzing, they can create input grammars or models based on their understanding of input formats, thereby improving the quality of generated test cases \cite{pham2019smart, aschermann2019nautilus, wang2019superion}. They can also write or modify test harnesses (a.k.a. fuzz drivers) to overcome reachability issues. With deeper insights into the SUT, they may annotate source code to improve the feedback signals used in coverage-guided fuzzing, enabling it to reach deeper program states \cite{aschermann2020ijon}.

Despite this, to the best of our knowledge, there has been limited research on developing techniques to \emph{assist humans} in analyzing fuzzing results and extracting actionable insights for improving fuzzing campaigns. Existing tools either support only simple applications \cite{shoshitaishvili2017rise} or offer basic visualizations (e.g., call graphs and control flow graphs) for manual analysis \cite{zhou2019visfuzz}. Although \fuzzintrospector{} recently introduced advanced features like fuzz blocker identification, developers and researchers must still manually analyze these blockers—making the process time-consuming and error-prone \cite{gao2023beyond}. Specifically, in~\cite{gao2023beyond}, we—particularly the first author—spent over 260 hours analyzing and classifying the top fuzz blockers reported by \fuzzintrospector{} in widely used projects such as LibPNG\footnote{\url{https://github.com/pnggroup/libpng}}, iGraph\footnote{\url{https://github.com/igraph/igraph}}, and OpenSSL\footnote{\url{https://github.com/openssl/openssl}}. While this effort was worthwhile—it produced a classification of fuzz blockers with clear characteristics and led to unblocking 10 fuzz blockers in LibPNG and boosting code coverage from 39.29\% to 51.67\%\footnote{\url{https://github.com/pnggroup/libpng/pull/551}}—it is largely impractical for most developers and researchers to invest such substantial effort.


Hence, a natural question arises: \emph{Can we automate the process of fuzz blocker analysis and provide analysts with a more intuitive interface to extract insights from a fuzzing campaign?} To address this question, we design the first human-assisting framework for analyzing fuzzing results. It leverages our fuzz blocker classification from \cite{gao2023beyond} and incorporates both pattern-matching and analytical techniques to identify types of fuzz blockers and suggest possible solutions to unblock them. We establish the following criteria for our design:



\begin{itemize}
        \item \textbf{C1 - Functionality.} The framework should offer a \emph{holistic view} of the SUT, enabling humans to classify fuzz blockers and extract insights from both static code and runtime data gathered during a fuzzing campaign. For example, when analyzing a blocking branch condition, analysts should be able to trace controlling variables (e.g., via static backward slicing), verify whether they are tainted (e.g., via dynamic taint analysis), and inspect their runtime values to estimate if certain constraints are likely to be met.

        \item \textbf{C2 - Applicability \& Scalability.} The framework should support large, real-world systems.

    \item \textbf{C3 - Extensibility.} The framework should have a modular design to support new use cases and facilitate future integration with AI components. Since human time is valuable, we envision an interface where AI can learn from past human input to automatically address simpler blockers.

    \item \textbf{C4 - Usability.} The framework should integrate smoothly with common development tools (e.g., Microsoft Visual Studio Code) to enable intuitive and efficient analysis.
\end{itemize}

To implement this design, we developed a framework called \insightql{}, which consists of (i) a unified code database combining static and dynamic information (addressing C1),  
(ii) a parameterized query interface (addressing C3), and  
(iii) several VS Code plugins to enhance usability (addressing C4). The code database and query interface are built on top of the state-of-the-practice static analysis tool \codeql{}, ensuring strong applicability and scalability (C2). Our VS Code plugins allow users to conduct several analyses intuitively by simply right-clicking to select the statement or variable of interest and invoking the appropriate queries. We discuss these components in details in Sections~\ref{sec:codedb} and~\ref{subsec:query}, respectively.

In this paper we make the following contributions:

\begin{itemize}
    \item We present \insightql{}, the first integrated human-assisting framework for analyzing and solving fuzz blockers. It features a unified database that offers a holistic view of the System Under Test and a user-friendly, parameterized query interface.
    \item \insightql{} significantly assisted in analyzing and unblocking fuzz blockers in 14 popular real-world libraries, resulting in substantial code coverage improvements (up to 13.90\%).
\end{itemize}

We will release \insightql{} as an open-source framework upon acceptance at \color{blue}\url{https://github.com/melbournefuzzinghub/insightql}\color{black}~to foster future research on human-assisted fuzzing in general and human-assisting tooling in particular.


%

\section{Background and Motivating Example} \label{sec:overview}


In this section, we introduce the fundamentals of human-assisted fuzzing, present a classification of fuzz blockers, and provide a motivating example to illustrate how our tool, \insightql{}, assists analysts in quickly identifying and resolving complex fuzz blockers—thereby improving the efficiency and effectiveness of fuzz testing.

\begin{figure}[h]
  \centering
  \includegraphics[width=\linewidth]{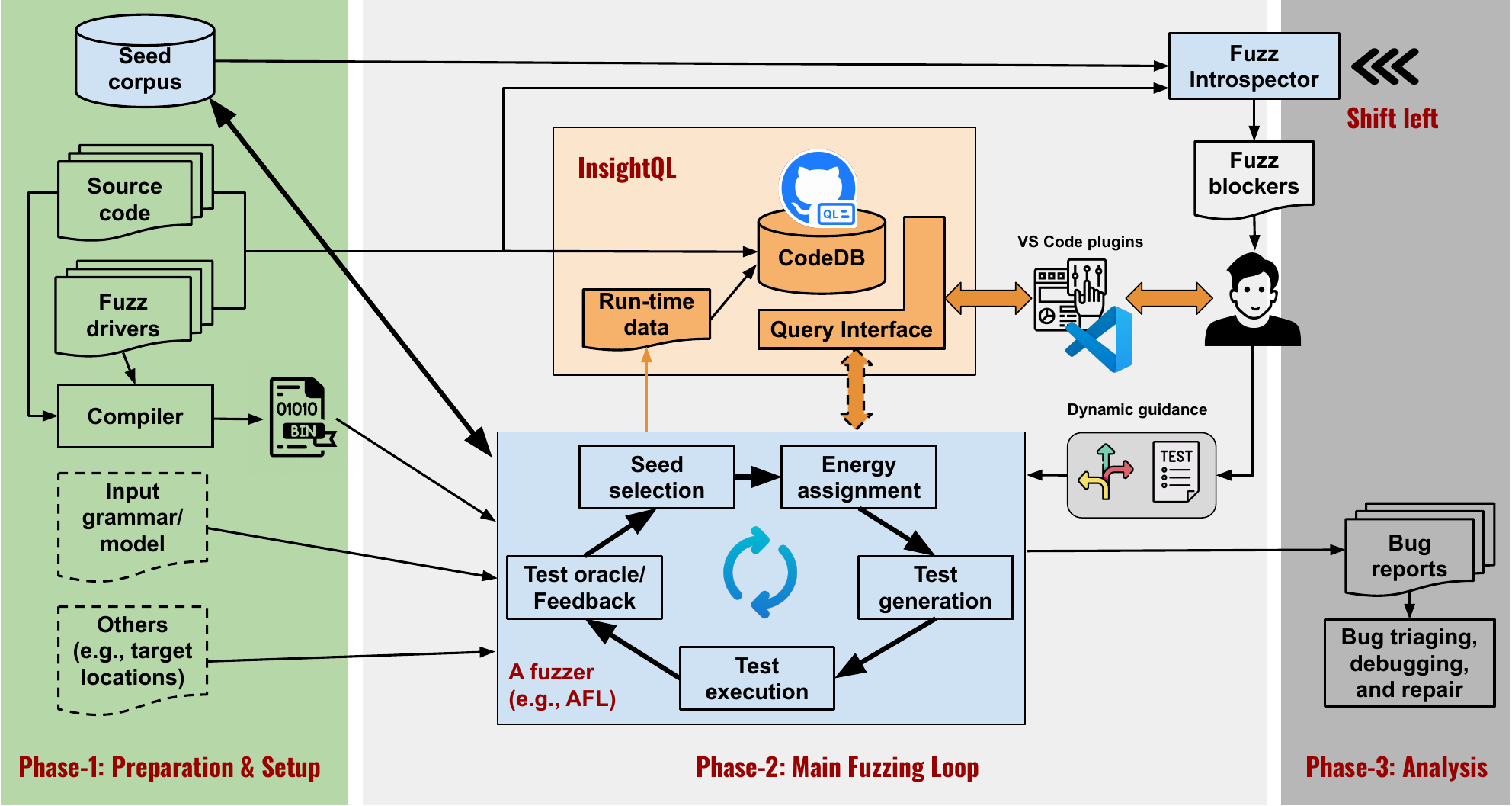}
  \caption{A 3-phase workflow of using code coverage-guided greybox fuzzing for vulnerability discovery} \label{fig:hitl}
\end{figure}

\subsection{Introduction to Human-Assisted Fuzzing} \label{subsec:fuzzingoverview}

Fuzzing is an automated process that repeatedly and intelligently generates ``random'' inputs (i.e., test cases) and feeds them to the system under test (SUT) to maximize code coverage and uncover bugs \cite{manes2019art}. A fuzzer is typically classified as either mutation-based or generation-based, depending on whether it produces test inputs from existing inputs or from scratch, respectively—and as blackbox, whitebox, or greybox based on the level of information it accesses from the SUT. Among these, greybox mutation-based fuzzing is arguably the most scalable and effective approach today. Unlike blackbox fuzzing (which blindly generates inputs) or whitebox fuzzing (which relies on detailed knowledge of the SUT), greybox fuzzing strikes a balance by using lightweight feedback from the SUT—often tracking which branches or lines each input covers—to guide its search algorithm. Several open-source coverage-guided fuzzing (CGF) tools implement this method, including AFL/AFL++ \cite{fioraldi2020afl++}, libFuzzer \cite{libfuzzer}, and Honggfuzz \cite{honggfuzz}.

Figure~\ref{fig:hitl} illustrates a typical workflow for using coverage-guided fuzzing (CGF) to discover vulnerabilities. This process is typically divided into three phases: Preparation \& Setup, the Main Fuzzing Loop, and Analysis, with varying degrees of human involvement. In Phase 1 (Preparation \& Setup), given the source code of the SUT and associated fuzz drivers, a human (e.g., developer, tester, or security researcher) compiles the code into a binary. During compilation, instrumentation is applied to enable efficient code coverage tracking and broader bug detection (e.g., using sanitizers like AddressSanitizer and MemorySanitizer). Since CGF primarily relies on mutation, the human also prepares a seed corpus containing sample inputs, as research has shown that the quality of the seed corpus significantly impacts fuzzing effectiveness \cite{herrera2021seed}. With recent advancements, the human may also optionally provide an input grammar or model to enable structure-aware fuzzing \cite{pham2019smart, aschermann2019nautilus, wang2019superion}, or specify targets for directed fuzzing \cite{bohme2017directed}.

In Phase 2 (Main Fuzzing Loop), during each iteration, a CGF fuzzer like AFL will (1) select a sample input from the corpus, (2) assign suitable fuzzing energy (i.e., how many mutants to generate), (3) generate new inputs, (4) execute them against the program under test (PUT), and (5) observe the PUT’s behavior. If a newly generated input triggers new behavior (e.g., a new branch is covered), the fuzzer adds it to the seed corpus for further fuzzing. If abnormal behavior is detected (e.g., crashes, memory errors), the input is saved and a report is generated for later analysis in Phase 3. This loop continues until a timeout is reached or the user stops the process. \emph{In traditional fuzzing, there is little to no human involvement during this phase.}

In Phase 3 (Analysis), human analysts review bug reports, triage issues, debug them, and develop patches. They also analyze logs and outputs (e.g., from FuzzIntrospector) to identify fuzz blockers \cite{gao2023beyond}, diagnose root causes, and suggest improvements—such as adding seeds, updating targets, or modifying fuzz drivers. Traditionally, this feedback is incorporated in later rounds of fuzzing. However, this delayed feedback can be inefficient, with fuzzers potentially running unproductively for long periods. Recent work \cite{bohme2020fuzzing, fang2024ddgf} has begun to move fuzz blocker analysis and resolution into Phase 2 (i.e., shifting left), allowing human intervention while fuzzing is in progress. Nonetheless, much of the difficult analysis remains manual, with minimal tool support.



\subsection{Fuzz Blocker Classification}
\label{subsec:classification}

\begin{figure}[ht]
\centering
\scalebox{0.8}{ 
\begin{forest}
for tree={
    draw,
    rounded corners,
    align=center,
    parent anchor=east,
    child anchor=west,
    grow'=0,
    edge={-},
    minimum width=2.8cm,
    minimum height=0.8cm,
    l sep=6pt,
    s sep=6pt,
    font=\small,
}
[?, minimum width=0.5cm
  [1.Tainted, minimum width=1.5cm
    [1.1.Magic Numbers]
    [1.2.Missing Extreme Inputs]
  ]
  [2.Not Tainted, minimum width=1.5cm
    [2.1.Missing Argument Options]
    [2.2.Missing Function Calls
      [2.2.1.Overloaded Functions]
      [2.2.2.Repeated Calls]
      [2.2.3.Config/Setting Calls]
      [2.2.4.Feature-Enabling Calls]
    ]
    [2.3.Wrong Call Order]
    [2.4.Unreachable Code]
  ]
]
\end{forest}
}
\caption{An in-progress classification of fuzz blockers \cite{gao2023beyond}}
\label{fig:blocker-taxonomy}
\end{figure}
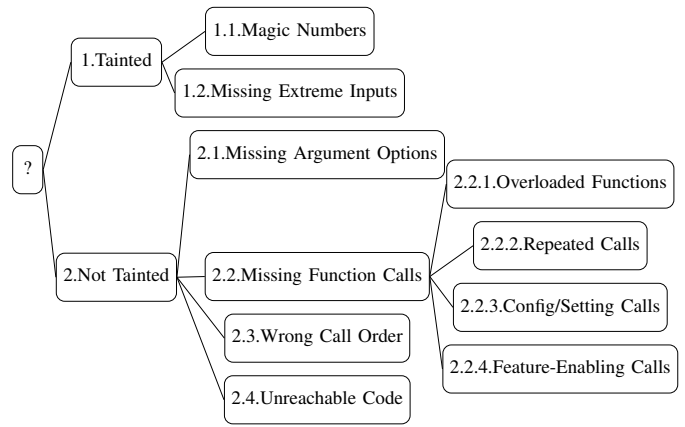

We believe that human analysts—and future AI-based solutions—can analyze and resolve fuzz blockers more effectively and efficiently when equipped with a clear taxonomy of fuzz blocker types and their characteristics. This insight motivated our work in \cite{gao2023beyond}. In our preliminary study, we manually analyzed three popular projects (LibPNG, igraph, and OpenSSL) and produced a classification of fuzz blockers into distinct types and sub-types, as illustrated in Figure~\ref{fig:blocker-taxonomy}. At the top level, we distinguish between input-dependent (i.e., tainted) and input-independent (i.e., not tainted) fuzz blockers. Input-dependent blockers are further divided into categories such as those caused by the use of magic numbers or missing ``extreme" inputs—for example, graph libraries requiring extremely large graphs to trigger certain code paths.

In contrast, input-independent blockers primarily arise from limitations in the fuzz driver itself. We categorize these into four subtypes: (2.1) missing valid options for function arguments (e.g., omitted values in enumerated types), (2.2) missing function calls, (2.3) unexplored or incorrect ordering of function calls, and (2.4) unreachable code due to external configurations. Among these, blockers caused by missing function calls can be further classified into several cases: (2.2.1) missing calls to overloaded functions (i.e., functions with the same name but different signatures), (2.2.2) missing repeated calls, (2.2.3) missing calls for setup or configuration, and (2.2.4) missing feature-enabling calls (e.g., enabling image transformations in the LibPNG library). Notably, our analysis revealed that the majority of top fuzz blockers in the studied libraries are input-independent. This suggests that simply increasing fuzzing time is often insufficient; instead, uncovering and resolving these blockers typically requires targeted modifications to the fuzz driver.

In \insightql{}, we provide an intuitive query interface together with a set of predefined and extensible queries—both powered by a unified code database—that help analysts identify fuzz blockers according to our classification. The process typically begins with a taint-analysis query to determine whether a fuzz blocker is input-dependent, and then proceeds with additional queries to identify sub-types, following the classification in Figure~\ref{fig:blocker-taxonomy} from left to right. Moreover, \insightql{} can offer guidance on how to resolve certain blockers. For example, it can suggest specific function calls that are missing from the fuzz driver and should be added to unblock the corresponding execution paths. We demonstrate this in the example below.

\subsection{Motivating Example} \label{subsec:example}


Listing \ref{lst:example} presents a sample libFuzzer-style fuzz target designed to challenge CGF fuzzers. This example simulates several operations from the real-world LibPNG library, including parsing data from the input buffer (line 28), setting options (line 29), and performing various transformations based on the configured flags (lines 30–33). It contains two fuzz blockers at lines 16 and 33, which prevent the execution of the code at lines 17 and 34, respectively.

The first fuzz blocker (line 16) is input-dependent, simulating CVE-2018-13785, a division-by-zero bug in LibPNG. To trigger the synthetic bug at line 17, specific values for \texttt{x}, \texttt{y}, and \texttt{z} data fields of the Image structure pointed by imgp are required—\texttt{x} must be 0x55555555, \texttt{y} must be 3, and \texttt{z} must be 1. Given the constraints on these values (e.g., based on the file format specification), generating an input that simultaneously satisfies such ``magic numbers" is highly challenging for a CGF fuzzer and is considered ``The Achilles' Heel of mutation-based fuzzing” \cite{hazimeh2020magma}.

The second fuzz blocker (line 33) is input-independent; the guarding condition cannot be satisfied by the given fuzz target, regardless of the data buffer's content. The only way to remove this blocker is by adding a call to the \texttt{set\_option\_B} function, for instance, after calling \texttt{set\_option\_A} (line 29).

While this simplified example is easy to analyze for demonstration purposes, identifying similar fuzz blockers in real-world libraries—often comprising thousands of lines of code across multiple files—is substantially more difficult. Without a clear understanding of the root cause and type of a fuzz blocker, developers may resort to generic strategies such as extending the fuzzing duration, switching to fuzzers that track or infer input-to-code relationships (e.g., RedQueen \cite{aschermann2019redqueen}, Greyone \cite{gan2020greyone}), or adopting alternative techniques like concolic or symbolic execution—\emph{often relying on trial and error.}

\begin{lstlisting}[caption={A motivating example with two fuzz blockers}, label={lst:example}, language=c]
#define OPTION_A 0x01
#define OPTION_B 0x10
typedef struct {
  uint8_t x;
  uint32_t y; //y must be between 1 and 4
  uint8_t z; //z must be 1 or 2
  //other data fields
} Image; 
int32_t flags = 0x11111100;
Image* parse(const uint8_t *data) {
  //parse data into an Imgage struct
}
void process(Image* imgp) {
  //do something
  uint32_t t = imgp->x * imgp->y * imgp->z;
  if (t == 0xFFFFFFFF) { //fuzz blocker 2
    abort();
  }
}
void set_option_A(){ 
  flags |= OPTION_A; 
} 
void set_option_B(){ 
  flags |= OPTION_B; 
} 
int LLVMFuzzerTestOneInput(const uint8_t *data, size_t size) { 
  if (size < 8) return 1;
  Image* img = parse(data);
  set_option_A();
  if ((flags & OPTION_A) != 0) { 
    //do some transformation
  }
  if ((flags & OPTION_B) != 0) { //fuzz blocker 1
    //do other transformation    
  }
  //do something else before processing
  process(img);
  return 0;  
} 
\end{lstlisting}




With \insightql{}, this process becomes significantly more efficient. Specifically, to analyze the first fuzz blocker, the developer opens the code in VS Code and right-clicks on the variable \texttt{t} (line 16) and chooses to run the hybrid taint-analysis query (Query-1 in Listing~\ref{lst:main-query-1}, Section~\ref{sec:approach}). \insightql{} traces the taint flow from the input buffer (\texttt{data}) to the \texttt{img} pointer (line 28), through the fields of the \texttt{Image} struct, and finally to \texttt{t}. It then suggests that \texttt{t} is tainted, indicating that this blocker is input-dependent. Existing static taint analysis engines such as \codeql{} often suffer from
under-tainting and fail to detect this dependency, as we will discuss in Section
3. The developer can then run another query (Query-2 in Listing~\ref{lst:main-query-2}, Section~\ref{sec:approach}) to examine the runtime value distribution of \texttt{t}, which reveals that values near zero are rare. This insight may suggest adopting concolic execution or enabling advanced fuzzer features such as AFL++’s CmpLog or libFuzzer’s value profiling to better guide input generation.

For the second blocker, the developer again right-clicks on the variables \texttt{flags} and \texttt{OPTION\_B} and runs the taint-analysis query to check whether they are tainted. In this case, neither variable is tainted, so \insightql{} determines that the fuzz blocker is input-independent. Following the classification in Figure~\ref{fig:blocker-taxonomy}, the developer then right-clicks on \texttt{OPTION\_B} (line 33) to investigate whether it is unset due to a missing function call (Query-3 shown in Listing~\ref{lst:main-query-3}, Section~\ref{sec:approach}). \insightql{} performs a data flow analysis and confirms that the condition indeed depends on a call to \texttt{set\_option\_B()}, which is absent. Based on this insight, the developer can update the fuzz driver—for example, by inserting a call to \texttt{set\_option\_B()} alongside the existing call to \texttt{set\_option\_A()} at line 29—to resolve the blocker.

\section{\insightql{}: Architecture and Design} \label{sec:approach}


\begin{figure}[h]
  \centering
  \includegraphics[width=1.0\linewidth]{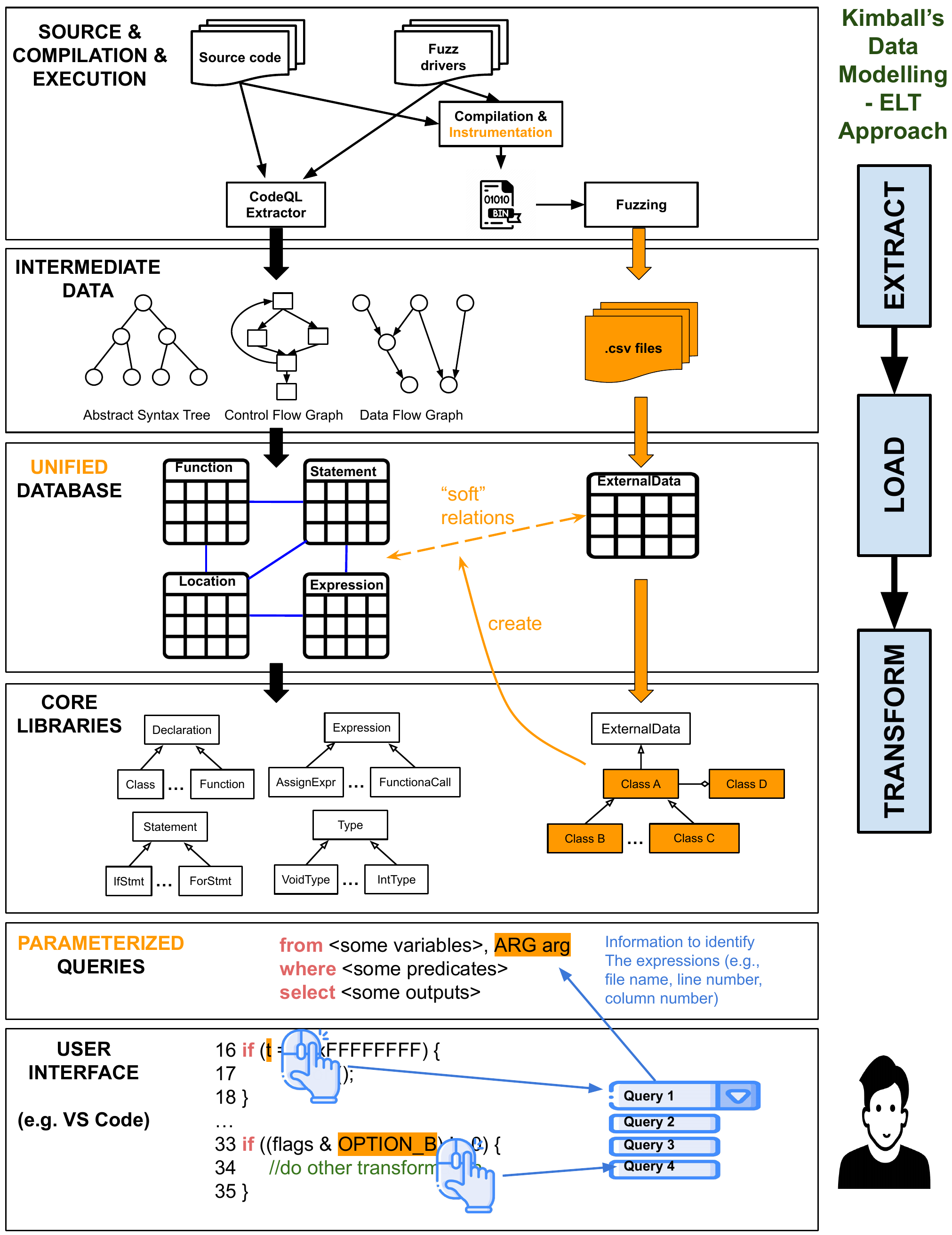}
  \caption{Multi-layer design of \insightql{}. \insightql{} is built on top of \codeql{}, following the Kimball's Data Warehouse model. New components are highlighted in orange.} \label{fig:architecture}
\end{figure}

Figure~\ref{fig:architecture} illustrates the complete architecture of \insightql{}, encompassing both back-end and front-end components designed to help users analyze fuzz blockers. The architecture can be viewed in multiple layers. From a user’s perspective, the primary focus is on the User Interface (bottom layer), which integrates seamlessly with Visual Studio Code. Users can simply right-click on expressions of interest, choose a suitable query to run, and receive results that assist in resolving fuzz blockers, as demonstrated in the motivating example. Achieving this required systematic work on data modeling to establish a unified database, as well as the development of a parameterized query interface to simplify user tasks. The following sections first propose the human-assisted fuzzing workflow and describe these two critical components in detail.

\subsection{Data Modeling with Kimball's Data Warehouse Approach} \label{sec:codedb}

We chose to build a code database for \insightql{} using \codeql{} instead of other static application security testing (SAST) tools like Infer and Joern because CodeQL offers key advantages, including its intuitive query language and database-driven analysis. While tools like Joern also support custom queries, CodeQL’s SQL-like syntax and object-oriented programming model make it more accessible and easier to write and understand. This enables more precise and customizable detection of complex, project-specific vulnerabilities. With additional support for multiple programming languages and seamless integration into CI/CD pipelines, CodeQL proves to be a powerful and flexible tool for secure code analysis.

With \codeql{} as the source of static information—ensuring high applicability, scalability, and extensibility—we needed to choose a suitable data modeling approach to effectively integrate the dynamic information generated by fuzzing into \codeql{}’s existing code database. This integration results in a unified database that offers a holistic view of the entire system under test.

\subsubsection{Introduction to CodeQL: Code as a Database}

CodeQL is a powerful static analysis tool originally developed by Semmle, a company later acquired by GitHub in 2019. It builds on over a decade of research by a team from Oxford University. CodeQL is particularly effective for automatically scanning applications for vulnerabilities—especially for variant analysis—and it also supports manual code review.

The core concept behind CodeQL is treating code as data. It generates a relational database that captures both syntactic features (e.g., the Abstract Syntax Tree or AST) and semantic features (e.g., the Control Flow Graph (CFG) and Data Flow Graph (DFG)) of a software project. This database generation process is illustrated on the left side of the top three layers in Figure~\ref{fig:architecture}. The resulting code database includes numerous tables—such as \texttt{Function}, \texttt{Statement}, \texttt{Expression}, and \texttt{Location}—which represent elements from the AST, CFG, and DFG. For example, a function may contain multiple statements, each mapped to specific source code locations.



For compiled languages like C and C++, CodeQL generates this relational database during the build process using a language-specific schema. Although GitHub provides open-source access to the query interface and a collection of reusable queries, it keeps the database generation component proprietary. This makes it difficult to modify the schema or introduce new tables and relationships. Currently, no publicly available specification exists for the schema syntax.\footnote{\url{https://codeql.github.com/docs/codeql-overview/codeql-glossary}} \emph{Given this, we must explore data modeling and integration strategies that avoid modifying CodeQL’s underlying database schema.}

\subsubsection{Data Modeling with Kimball's Data Warehouse Approach}

We apply the data warehouse model to combine static and dynamic information, which is an approach widely used to integrate heterogeneous data sources for efficient analysis \cite{inmon2005building, vaisman2014data}. Data warehouses store both current and historical data in a format that facilitates reporting, querying, and insight generation, aligning well with the goals of \insightql{}.

There are two widely adopted approaches to building a data warehouse: Inmon’s and Kimball’s models \cite{yessad2016comparative}. The primary difference between them lies in their data organization strategies. Inmon's model uses a normalized relational design, restructuring all data sources into a centralized relational schema. However, this approach is not suitable for our needs due to the proprietary constraints on modifying CodeQL’s database schema, as well as the substantial upfront effort required for data cleaning and restructuring. In contrast, Kimball’s model uses a dimensional design, where data is organized into fact and dimension tables. This approach typically adopts a star schema: a central fact table (representing measurable events) is surrounded by dimension tables (containing descriptive attributes), forming a data mart--a focused unit of analysis. Additional analyses can be supported by reusing shared dimension tables across data marts \cite{chaudhuri1997overview}.

To illustrate, consider a Point-of-Sale system in a retail store. A fact table can be created from order transactions, while dimension tables—such as \texttt{Products} and \texttt{Promotions}—surround the fact table to form a star schema.\footnote{\url{https://www.holistics.io/books/setup-analytics/kimball-s-dimensional-data-modeling/}} These tables are connected via foreign keys; for instance, each order may reference multiple products and applicable promotions. Over time, the fact table grows rapidly (e.g., hundreds of daily orders), whereas the dimension tables remain relatively small and are updated less frequently.

Following Kimball’s model, tables in \insightql{} containing static information serve as dimension tables, while dynamic, fuzzer-generated data is modeled as fact tables. To implement this approach, we must overcome two key challenges: (i) how to link fact tables with dimension tables without modifying CodeQL’s underlying schema, and (ii) how to define the structure of the fact tables. We address these challenges in Sections~\ref{subsec:elt} and~\ref{subsec:dynamic}, respectively.

\subsubsection{Connecting fact tables to dimension tables} \label{subsec:elt}


Merging different data sources into a Kimball-style data warehouse is commonly achieved through the Extract–Transform–Load (ETL) workflow. In ETL, data is first extracted from its sources, then transformed into the warehouse schema, and finally loaded into the target system. On the other hand, \insightql{} adopts the Extract–Load–Transform (ELT) approach \cite{haryono2020comparison}, a modern variation in which raw data is extracted and loaded directly into the warehouse, with transformations applied afterwards using queries. 

Figure~\ref{fig:architecture}—particularly the first four layers—illustrates how \insightql{} builds a Kimball-style data warehouse using the ELT approach. In the extraction step, in addition to the static code database generated by \codeql{}, \insightql{} instruments the fuzzing binary to collect runtime values (e.g., selected variables), which are stored in CSV files. Section~\ref{subsec:dynamic} details the types of dynamic data collected during fuzzing, chosen to align with Kimball’s guidelines for optimizing performance. \insightql{} then loads these dynamic data files into CodeQL’s built-in \texttt{ExternalData} table type, treating them as fact tables. These tables remain logically separate from the existing dimension tables, which store static information. The transformation step is deferred and supports task-specific analysis. Under the ETL variant, the transform stage would pre-compute joins between fact and dimension tables and other operations (e.g., normalization, aggregation) before loading; as a result, answering new analytical questions would typically require modifying and re-running upstream pipelines.


\subsubsection{Defining the structure of dynamic data} \label{subsec:dynamic}

According to Kimball’s data warehouse guidelines \cite{kimball2011data}, one must ``decide on the grain'' of the fact tables—that is, determine the level of detail at which data will be stored. This level of detail should be as fine-grained (or atomic) as possible, meaning it cannot be further subdivided. For example, in the earlier Point of Sales scenario, the appropriate grain would be each line item within an order, rather than the order as a whole. This level of detail allows business users to answer questions like, “Which products sold the most in our stores today?”, which would not be possible using only order-level data.

The granularity is typically dictated by the most detailed level required to answer relevant analytical questions. In the case of \insightql{}, we define the grain along two specific dimensions: \emph{code location} and \emph{execution time}.

For code location, we adopt \emph{variable-level granularity}, which means that we track the value of each variable at runtime. For execution time, we use \emph{basic block-level granularity}, where we measure the number of times each basic block is executed during a fuzzing run. Together, these dimensions allow us to uniquely record the value of a given variable each time a specific basic block is executed. This design enables queries such as: “What is the value of the variable \texttt{t} at the branch \texttt{if (t == 0xFFFFFFFF)} when the total number of basic blocks executed reaches 1000?” Such insights are crucial for understanding how input data propagates through the program and for identifying fuzz blockers that depend on complex value combinations at specific control-flow locations.

The Kimbal data warehouse design provides two main benefits. First, by adopting the ELT workflow, the warehouse can reuse the same dynamic data across many analytical tasks, with task-specific transformations applied as needed. Second, by defining the grain of fact tables along code location and execution time, it supports tracking the history of fuzzing runs and comparing human-assistance strategies across time and code locations.

\subsection{ \insightql{} query interface and parameterized queries}\label{subsec:query}

\subsubsection{Introduction to CodeQL query interface}

To extract information from the constructed code database, \codeql{} provides a query language called QL. The syntax of QL is similar to SQL, but its semantics are based on Datalog—a declarative logic programming language commonly used for querying. As a result, QL is fundamentally logic-based, and all operations in QL are expressed as logical operations\footnote{\url{https://codeql.github.com/docs/ql-language-reference/about-the-ql-language/}}. CodeQL adopts the object-oriented programming paradigm and includes a set of core libraries to enhance usability. These libraries consist of numerous classes, each representing data from a single table or a join of multiple tables in the database. From a high-level perspective, a QL query uses set operations to filter out uninteresting information and extract relevant code patterns.

\begin{lstlisting}[language=sql, caption=Sample query using from-where-select, breaklines=true, label=lst:query1]
from FunctionCall fc, Function callee, Function main
where main.getName() = "main" 
and fc.getEnclosingFunction() = main 
and fc.getTarget() = callee 
select callee
\end{lstlisting}

\begin{lstlisting}[
    language=sql,
    caption={Sample query using predicates},
    label={lst:query2},
    breaklines=true,
    morekeywords={predicate,class,extends}
]
predicate isFunctionCallInMain(FunctionCall fc, Function main){
  main.getName() = "main" 
  and fc.getEnclosingFunction() = main
}
from  FunctionCall fc, Function callee, Function main
where isFunctionCallInMain(fc, main)
  and fc.getTarget() = callee
select callee
\end{lstlisting}

\begin{lstlisting}[
    language=sql,
    caption={Sample query using classes},
    label={lst:query3},
    breaklines=true,
    morekeywords={predicate,class,extends}]
class FuncCallInMain extends FunctionCall {
  FuncCallInMain() {
    exists (Function main |
      this.getEnclosingFunction() = main |
      main.getName() = "main")
  }
}
from  FuncCallInMain fcMain, Function callee
where fcMain.getTarget() = callee
select callee
\end{lstlisting}

Developers can use the same approach to write custom queries, each tailored to detect a specific pattern of interest (e.g., identifying potential use-after-free vulnerabilities)\footnote{\url{https://github.com/githubsatelliteworkshops/codeql-cpp/blob/main/workshop.md}}.  Before presenting the extended queries introduced by \insightql{}, we first provide a few simple examples and introduce key concepts in the QL language, such as \texttt{class} and \texttt{predicate}.

Listings~\ref{lst:query1}, \ref{lst:query2}, and \ref{lst:query3} present three \codeql{} queries that accomplish the same simple yet useful task: identifying all functions called by a function named \texttt{main}, using three different approaches. These approaches demonstrate the use of (i) basic SQL-like syntax, (ii) a predicate (i.e., a reusable logic unit), and (iii) a class-based structure. In each query, the \texttt{from} clause declares the variables to be used. These variables can be instances of types defined in the core libraries or custom classes. The \texttt{where} clause specifies the logical conditions that the declared variables must satisfy, while the \texttt{select} clause determines what information is returned for those variables.

\codeql{} supports predicates and classes to improve the modularity, reusability, and composability of queries. A QL predicate is a self-contained logical unit, essentially a small \texttt{from-where-select} query, that encapsulates logic and can be reused across multiple queries. This makes predicates especially useful for abstraction and debugging. QL classes, on the other hand, allow developers to define new types that represent program elements matching specific patterns. These can be thought of as virtual tables and can be used in queries just like built-in types. Following the object-oriented paradigm, QL classes can inherit members (i.e., predicates and methods) from parent classes. Custom predicates defined in these classes are particularly useful for joining virtual tables and filtering data more expressively.



\subsubsection{Parameterized queries}

In \insightql{}, we extend the query set to enable the analysis of static information, dynamic information, or a combination of both. In this section, we explain what a parameterized query is, and present three parameterized exemplary queries. 

Traditional \codeql{} queries follow a top-down approach, scanning the codebase from modules down to functions and statements-effective for vulnerability detection or pattern-based reviews. However, fuzz blocker analysis typically requires a bottom-up approach: starting from identified blockers (e.g., those reported by FuzzIntrospector~\cite{fuzzintrospector}) and tracing backward to find related data sources, control predicates, and functions, as illustrated in our motivating example.

\begin{lstlisting}[language=sql, caption=A sample parameterized query, breaklines=true, label=lst:query-parameterized-example]
Class ARG {
 ARG() { this = "func_name@@file_name@@129@@9"}
 int getIntAt(int index) { ... }
}
from FunctionCall fc, ARG arg
where callee.getName() = arg.getStringAt(0)
and callee.getStartLine() = arg.getIntAt(1)
select callee
\end{lstlisting}

Accordingly, the query must accept blocker-specific context at invocation time. To support this, we introduce parameterized query templates that take minimal arguments, along with an automated argument synthesizer to populate them. 

To streamline this, \insightql{}’s VS Code plugin captures fuzz blocker's information when the developer right-clicks and automatically supplies it as a query argument. Internally, we use a simple context-free grammar to synthesize arguments in a structured format (e.g., ``\texttt{func\_name@@\allowbreak file\_name@@\allowbreak start\_line@@\allowbreak start\_column}
''). These arguments are encapsulated in a \codeql{} class, \texttt{ARG}, which provides member functions such as \texttt{getIntAt()} for indexed retrieval. The initialization of this class is dynamically produced by the argument generator. Within the query body, placeholders at Lines 6–7 in Listing~\ref{lst:query1}) are replaced with calls to \texttt{arg.getStringAt(0)} and \texttt{arg.getIntAt(1)}. This transformation effectively constrains the analysis to a blocker-specific search space and bounds the result set. 


Listing~\ref{lst:main-query-1} presents the hybrid taint-analysis query (Query-1) used to check whether the variable \texttt{t} in the first fuzz blocker from the motivating example is (likely) tainted. This blocker—adapted from CVE-2018-13785—poses a challenge for \codeql{}’s built-in taint analysis, which lacks support for pointer and alias analysis. Consequently, it fails to associate the \texttt{img} pointer (defined at line 28) with the \texttt{imgp} pointer in the \texttt{process} function and cannot trace the data flow from the input buffer \texttt{data} to \texttt{t}. \emph{Our insight is that if, at runtime, two pointers reference the same memory, this missing link can be inferred.}

We extend this insight to capture more general cases. Technically, given a group of variables called $S$ and another variable named $s'$ that is not in $S$, if there exists a function $f$ such that $s' = f(S)$, then $s'$ is likely dependent on $S$. If $S$ is tainted, we can infer that $s'$ is also likely tainted. In the motivating example, the relationship appears as $imgp = img$ (these two pointers point to the same memory area), but this can be generalized to linear (e.g., $imgp = img + 10$) or non-linear functions (e.g., $s' = y^2 + 2z + 1$; $y$ and $z$ are in $S$).

\begin{lstlisting}[language=sql, caption=InsightQL Query-1: check if a variable is (likely) tainted. It uses static data flow first and leverages dynamic data flow if necessary, label=lst:main-query-1]
predicate isTainted(source, sink){
    ( isStaticallyTainted(source, sink) )
    or
    (
      intermediateSources = forwardFlow(source)
      and intermediateSinks = backwardFlow(sink)
      and isLikelyTainted(intermediateSources, intermediateSinks) 
    )
}
from DataFlow::node source, DataFlow::node sink, ARG arg
where sink.getLocation() = arg.getLocation()
and isTainted(source, sink)
select source, sink 
\end{lstlisting}

\begin{algorithm}[h]
\caption{\textsc{isLikelyTainted}}
\label{alg:updatecallgraph}
\textbf{Input:} Project Source Code $P$, CodeQL Database $DB$, Intermediate Sources $S$, Intermediate Sinks $S'$ \\
\textbf{Output:} Likely Tainted $likelyTainted$
\begin{algorithmic}[1]
\State $likelyTainted \gets False$ 
\State \textcolor{olive}{Fuzzing binary} $b \gets$ \textsc{Instrument}($P, S, S'$)
\State \textcolor{olive}{Runtime data} $d \gets$ \textsc{FuzzAndExtract}($b$)
\State \textcolor{olive}{Fact table} $f \gets$ \textsc{Load}($d$)
\State \textcolor{olive}{\insightql{} Database} $DB' \gets$ \textsc{Transform}($DB, f$)
\ForAll{$s' \in S'$}
    \State \textcolor{olive}{Variable-Value pairs} $(K, V) \gets$ \textsc{Query}($DB', S, s'$)
    \State \textcolor{olive}{Relationship} $R \gets$ \textsc{InferRelationship}($K, V$)
    \If{$R \neq \emptyset$}
        \State $likelyTainted \gets True$
        \State \textbf{break}
    \EndIf
\EndFor
\State \Return $likelyTainted$
\end{algorithmic}
\end{algorithm}


Based on this idea, we design our query as follows: given a data source (the \texttt{data} buffer in the example) and a potential sink (variable \texttt{t}), which is not considered tainted if only static taint analysis is used, we first identify all variables that can be reached from the source using forward dataflow tracking and call these intermediate sources $S$. Similarly, we find all variables that can flow to the sink using backward dataflow tracking and call these intermediate sinks $S'$.

We then pass both sets to a predicate named isLikelyTainted, which is described in Algorithm 1. Inside this predicate, InsightQL first instruments the fuzzing binary to collect runtime data for all variables in $S$ and $S'$. It then follows the ELT model (see Section~\ref{sec:codedb}) to extract, load, and transform run-time data to construct a unified database. For each variable $s'$ in $S'$, we check whether an equivalence relationship can be inferred between it and some subset of $S$. If such a relationship exists, we consider that there is a tainted link between $S$ and $S'$ and therefore establish a connection from the original data buffer to the sink of interest. To infer such a relationship, we integrate DIG~\cite{nguyen2012using}, a tool for numerical invariant generation, into \insightql{}.
While this way of inferring dataflow is not foolproof~\cite{schneider:2000}, it provides useful insights to the analyst, which can be leveraged to effectively identify input-dependent fuzz blockers.


Listing~\ref{lst:main-query-2} presents a query for estimating the runtime value distribution of a variable or expression. As described in the motivating example, it can be used to visualize the distribution of the variable \texttt{t} in the first fuzz blocker, helping assess the likelihood that the condition \texttt{t == 0xFFFFFFFF} will be satisfied during extended fuzzing runs. This estimation is performed using Kernel Density Estimation~\cite{kde}. The parameterized query provides an argument string containing the expression schema (e.g., variable \texttt{t} is calculated from \texttt{x} * \texttt{y}) and source locations of relevant operands. Within the query, placeholders are replaced with accessor calls to resolve these operands and reconstruct the expression. Thus, allowing us to reconstruct calculations executed at run-time, but capture a detailed granularity when gathering traces. 

\begin{lstlisting}[language=sql, caption=InsightQL Query-2: calculate distribution based on runtime values of numeric variables, label=lst:main-query-2]
arg = {"var1 * var2"@@func@@line@@col@@func@@line@@col}
from VariableAccess var1, VariableAccess var2, ARG arg
where var1.getLocation() = arg.getIntAt(2) and
var1.getLocation() = arg.getIntAt(5)
select distribution(var1 * var2)
\end{lstlisting}

Listing~\ref{lst:main-query-3} shows a query that addresses the input-independent fuzz blocker in the motivating example. We define two custom classes: \texttt{FlagSet} and \texttt{FlagUnSet}. The \texttt{FlagSet} class checks for a static dataflow from a function (called by the fuzz driver) to the blocker. Conversely, \texttt{FlagUnSet} identifies dataflows from functions that are not called. Both classes include member predicates to return the flags they control.

\begin{lstlisting}[language=sql, caption=InsightQL Query-3: detects the missing function call for fuzz blocker 1 in the motivating example, label=lst:main-query-3]
from FlagSet fs, FlagUnSet fu, Literal lt, ARG arg
where fs.getFlagAsBit() != fu.getFlagAsBit()
and lt.getLocation() = arg.getLocation()
and fu.getFlagAsBit() = lt.getValue().toBit()
select fu, fu.getEnclosingFunctions()
\end{lstlisting}

When the user right-clicks on \texttt{OPTION\_B} and runs this query, it determines whether the flag is unset and, if so, returns \texttt{set\_option\_B} as the missing function call. This is realized in three steps: (1) we call the member function \texttt{getFlagAsBit()} to identify flags in \texttt{FlagUnSet} not present in \texttt{FlagSet}; (2) we associate a parameterized \texttt{ARG} type with a literal that matches \texttt{OPTION\_B}’s location in the code; (3) we match that literal to the flag’s bit value. The \texttt{Literal} type is a built-in \codeql{} class representing macros.

Due to space constraints, we omit detailed implementations of the classes and predicates used in these queries. Complete code is available in our replication package.

\section{Implementation}\label{sec:implementation}

On the user interface side, we extended the official \codeql{} plugin for Visual Studio Code to support human-assisted fuzzing more effectively. Our custom extension introduces a “right-click to query” interaction model, allowing users to seamlessly trigger backend tasks such as database construction and fuzzer compilation directly from the IDE. 

On the back end, we implemented approximately 480 lines of \codeql{} code to support the three queries presented in Listings~\ref{lst:main-query-1}–\ref{lst:main-query-3}. To support taint-analysis query (Listing~\ref{lst:main-query-1}), we added two additional taint propagation rules, as recommended by CodeQL's documentation.\footnote{\url{https://codeql.github.com/docs/codeql-language-guides/advanced-dataflow-scenarios-cpp/}} These rules are: (1) pointer address copy (e.g., \texttt{ptr = (char*) input}; if \texttt{input} is tainted, then \texttt{ptr} is also considered tainted), and (2) field-to-qualifier propagation (e.g., \texttt{ptr->field = x}; if \texttt{x} is tainted, we consider \texttt{ptr} to be tainted).

These three queries are capable of analyzing six out of the nine fuzz blocker types (counted as leaf nodes in the classification in Figure~\ref{fig:blocker-taxonomy}), representing 66.67\% coverage. Specifically, they support blocker types 1.1, 1.2, 2.1, 2.2.1, 2.2.3, and 2.2.4. Currently, \insightql{} does not support blockers 2.2.2 (Repeated Calls) and 2.3 (Incorrect Call Ordering) due to the lack of semantic reasoning required to capture their behavior. These blockers were heuristically classified based on syntactic cues, for example, when the blocked code yields an error message stating, ``Do not call this function repeatedly." For blocker type 2.4 (Unreachable Code due to External Configuration), support is also lacking because it requires knowledge of build-time configuration that is not available in the codebase. 

It is important to note that the classification presented in Figure~\ref{fig:blocker-taxonomy} is currently incomplete, as our broader study of fuzz blockers~\cite{gao2023beyond} is still ongoing. As this work progresses, we anticipate expanding the taxonomy and augmenting it with additional queries. This creates a positive feedback loop: \insightql{} accelerates the analysis process by quickly identifying and classifying known types of blockers (see our evaluation), which in turn informs and refines the blocker taxonomy. As the classification becomes more comprehensive, we can implement new \insightql{} queries to detect additional patterns more effectively.

This iterative process mirrors the success of \codeql{}, on which \insightql{} is built. \codeql{}'s extensive set of queries has enabled researchers to uncover numerous vulnerabilities. When a new vulnerability type is discovered, researchers often contribute new queries to identify similar issues in other codebases. In fact, \codeql{} used to incentivize this through bounty programs that reward contributions to its query set~\footnote{\url{https://github.com/github/securitylab/discussions/828}}.

Moreover, we developed a Python-based framework consisting of roughly 6,500 lines of code to automate the core components of the \insightql{} workflow. This backend is responsible for constructing \insightql{} databases, compiling fuzzers, and extracting fuzz blockers for any C/C++ project integrated with OSS-Fuzz. It also supports complete end-to-end automation, including runtime data collection, fact table construction, generation of parameterized queries based on user interactions, and rendering of query results.


\section{Evaluation}\label{sec:evaluation_new}

In this section, we present the evaluation conducted to assess \insightql{} based on the criteria outlined in Section~\ref{sec:intro}. Since both \insightql{} and its base, \codeql{}, are extensible by design, we specifically evaluate their Functionality, Applicability \& Scalability, and Usability through the following research questions:

\begin{itemize}
\item \textbf{RQ-1.} Can \insightql{} be applied to a variety of real-world applications?
\item \textbf{RQ-2.} Is \insightql{} effective in helping users resolve fuzz blockers and improve code coverage?
\item \textbf{RQ-3.} Is \insightql{} easy to use?
\end{itemize}

\subsection{Evaluation Setup}

To answer RQ-1 and RQ-2, we applied \insightql{} to the popular FuzzBench benchmark~\cite{fuzzBench}. Among the 24 subject libraries in the benchmark, we excluded OpenSSL due to compilation errors with OSS-Fuzz, and 9 other libraries due to either compilation errors or incompatibility with FuzzIntrospector. For the remaining 14 eligible libraries, as shown in Table~\ref{tab:fuzz_blocker_stats}, we followed the steps below to conduct our experiments.

\begin{itemize}
    \item \textbf{Step-1.} Create a \codeql{} database for the subject.
    \item \textbf{Step-2.} Extract all fuzz blockers reported by \fuzzintrospector{}. We use the reports dated on 01/03/2025. 
    \item \textbf{Step-3.} For each fuzz blocker, invoke the appropriate query by right-clicking on the relevant variable or statement and selecting a query from a list. Currently, the list contains 3 queries as explained in Section~\ref{subsec:query}.
    \item \textbf{Step-4.} Use the information provided by the query to attempt to resolve the fuzz blocker.
    \item \textbf{Step-5.} Compile the modified fuzz driver and re-run fuzzing. Check whether the blocker is resolved and collect code coverage data to measure any improvement. If necessary, return to Step-4 for further refinement, or return to Step-3 to address other fuzz blockers.
\end{itemize}

To address RQ-3, we conducted a small-scale user study involving two co-authors, both PhD students with varying levels of experience in fuzzing and static analysis. The first participant (first author), who developed \insightql{}, has experience with both fuzzing and \codeql{}, and is an intermediate-level C/C++ programmer. The second participant (fourth author) has a basic understanding of C/C++ and fuzzing, but no prior experience with \codeql{}.

All experiments were carried out on an Ubuntu 22.04 workstation with 12 CPU cores and 64 GB of memory. We used libFuzzer as the underlying fuzzing engine, along with the default drivers and seed corpora from OSS-Fuzz.

\subsection{Results}

Table~\ref{tab:fuzz_blocker_stats} presents the 14 eligible projects used in our experiments (1\textsuperscript{st} column), their GitHub stars as a measure of popularity (2\textsuperscript{nd} column), the number of fuzz blockers reported by \fuzzintrospector{} (3\textsuperscript{rd} column), and the results of applying Query-1 (Listing~\ref{lst:main-query-1}), which targets input-dependent fuzz blockers, and Query-3 (Listing~\ref{lst:main-query-3}), which identifies input-independent fuzz blockers. We do not evaluate Query-2, as its results are more subjective and depend on interpretation of the value distributions as well as the invocation of additional tools such as concolic execution.

\begin{table*}[t]
  \centering
  \caption{Fuzz blockers identified and resolved by \insightql{} across eligible FuzzBench projects, together with line‐coverage comparison between the default OSS-Fuzz fuzz driver and the \insightql{}‐enhanced fuzz driver. }
  \label{tab:fuzz_blocker_stats}

  \begin{tabular}{lcc|cc|ccc}
    \toprule
    \textbf{Project} &
    \makecell{\textbf{GitHub}\\\textbf{stars}} &
    \makecell{\textbf{Fuzz}\\\textbf{blockers}} &
    \makecell{\textbf{Query-1}\\\textbf{statically tainted}} &
    \makecell{\textbf{Query-1}\\\textbf{likely tainted}} &
    \makecell{\textbf{Query-3}\\\textbf{applicable}} &
    \makecell{\textbf{Query-3}\\\textbf{solved}} &
    \makecell{\textbf{Coverage}\\\textbf{improvement}} \\ \midrule
    
    FreeType2     & 634  & 20  & 0  & 0                      & 3  & 3  & +10.17\% (+10,351) \\
    HarfBuzz      & 4.7k & 15 & 1  & 2                      & 3  & 3  & +3.30\%  (+1,929) \\
    Little-CMS    & 625  & 17  & 5  & 4                      & 8  & 5  & +3.99\%  (+681) \\
    LibPNG        & 1.4k & 23  & 3  & 0                      & 17 & 16 & +13.90\% (+1,786) \\
    libxml2       & 646  & 94  & 27 & 0                      & 0  & 0  & 0\% \\
    libxslt       & 65   & 7   & 0  & 4                      & 1  & 1  & +2.59\%  (+1,634) \\
    Mbed-TLS      & 5.9k & 35  & 19 & 7                 & 1  & 1  & +1.48\%  (+965) \\
    OpenH264      & 5.7k & 10  & 10 & 0               & 0  & 0  & 0\% \\
    OpenThread    & 3.7k & 26  & 3  & 4                 & 2  & 0  & 0\% \\
    RE2           & 9.3k & 15  & 0  & 10                & 0  & 0  & 0\% \\
    SQLite3       & 7.9k & 16  & 3  & 0                 & 0  & 0  & 0\% \\
    stb           & 28.9k& 7   & 0  & 0                      & 0  & 0  & 0\% \\
    zlib          & 6.2k & 3   & 0  & 0                      & 1  & 0  & 0\% \\ 
    Bloaty       & 5.1k & 7   & 0  & Instrumentation-Failed & Unknown  & Unknown  & Unknown \\ \midrule
    \textbf{Total}&      & 295 & 71 & 31                & 36 & 29 & \\ \bottomrule
  \end{tabular}
\end{table*}

\subsubsection{Answer to RQ-1 regarding the applicability of \insightql{}}

We evaluate whether \insightql{} can build a unified database containing both static and dynamic information and successfully run the predefined queries. As shown in Table~\ref{tab:fuzz_blocker_stats}, out of 14 eligible projects from FuzzBench, \insightql{} works on 13 projects, achieving an applicability rate of 92.86\%.

The only unsuccessful subject is Bloaty. \insightql{} was unable to instrument the code to collect runtime values, which prevented the construction of the unified database and support for Query-1—the prerequisite for running the other queries. This failure stems from difficulties injecting the commands required to invoke the \insightql{} instrumentation pass, likely due to the complexity of building Bloaty’s dependencies. We consider this an engineering issue that could be resolved with the involvement of Bloaty’s maintainers. In fact, complicated build systems have also compromised the applicability of several fuzzing-related projects, including \fuzzintrospector{}.

\textbox{\textbf{Answer to RQ-1:} \insightql{} works on 92.86\% of the 14 eligible real-world libraries from FuzzBench, demonstrating its high applicability and scalability.}

\subsubsection{Answers to RQ-2 regarding the effectiveness of \insightql{}}

In the 4\textsuperscript{th} column of Table~\ref{tab:fuzz_blocker_stats}, we report that 71 of the fuzz blockers (16\%) can be confirmed as statically tainted—that is, they can be identified using only the static data of the code database (see line 2 in Listing~\ref{lst:main-query-1}).

It is worth noting that due to limitations in \codeql{}’s static data flow analysis—particularly its conservative approach to balance between under-tainting and over-tainting—some forms of data flow are not explicitly supported. For instance, in Listing~\ref{lst:taint1}, \codeql{} fails to detect the seemingly straightforward taint flow from the data buffer to \texttt{user\_read\_data} (line 4). Similarly, in Listing~\ref{lst:taint2}, it cannot trace the taint flow from the data buffer to the \texttt{font} pointer (line 4), possibly due to the complex logic in functions such as \texttt{png\_set\_read\_fn} and \texttt{hb\_*\_create}.

\begin{lstlisting}[language=C, caption=Illustration of the fuzzing data entrypoint in LibPNG, label=lst:taint1]
png_handler.buf_state = new BufState();
png_handler.buf_state->data = data + kPngHeaderSize;
png_handler.buf_state->bytes_left = size - kPngHeaderSize;
png_set_read_fn(png_handler.png_ptr, png_handler.buf_state, user_read_data);
\end{lstlisting}

\begin{lstlisting}[language=C, caption=Illustration of the fuzzing data entrypoint in the Harfbuzz library, label=lst:taint2]
alloc_state = _fuzzing_alloc_state (data, size);
hb_blob_t *blob = hb_blob_create ((const char *)data, size, HB_MEMORY_MODE_READONLY, nullptr, nullptr);
hb_face_t *face = hb_face_create (blob, 0);
hb_font_t *font = hb_font_create (face);
\end{lstlisting}

In our experiments, we leveraged general fuzzing knowledge and basic code understanding to manually mark such variables as tainted, enabling Query-1 to follow longer taint propagation chains deeper into the fuzz blockers. \insightql{} needed that support for 49 out of 71 (69\%) statically tainted (input-dependent) blockers.

By applying Query-1 to further analyze potentially tainted fuzz blockers based on run-time data, \insightql{} confirmed 31 additional input-dependent fuzz blockers (as reported in the 5\textsuperscript{th} column), bringing the total number of confirmed input-dependent fuzz blockers to 102, accounting for 34.58\% of all studied fuzz blockers.

\begin{lstlisting}[language=C, caption=Illustration of two types of non-tainted fuzz blockers, label=lst:input-independent_example]
// root cause for sub-type 2.2.3: missing a configuration-related function call
void png_set_rgb_to_gray (png_ptr){
  png_ptr->transformations |= PNG_RGB_TO_GRAY
}
// root cause for sub-type 2.1: different hard-coded argument/options in the fuzz driver
image.format = PNG_FORMAT_RGBA; 
png_image_finish_read(&image, NULL, buffer.data(), 0, NULL);

// trigger for both sub-type 2.2.3 and sub-type 2.1: calling image transformation init
void png_init_read_transformation (png_ptr){  
  ...
  // fuzz blocker
  if ((png_ptr->transformation & PNG_RGB_TO_GRAY) != 0) { // unreached code  
    png_colorspace_set_rgb_coefficients(); }
}


\end{lstlisting}

While Query-1 enables taint analysis—which can be generally applied to check whether an arbitrary variable or object is tainted—Query-3 targets specific code patterns to detect input-independent fuzz blockers, particularly sub-types 2.1, 2.2.1, 2.2.3, and 2.2.4 in the classification shown in Figure~\ref{fig:blocker-taxonomy}. Listing~\ref{lst:input-independent_example} presents a concrete example from the LibPNG library, which exhibits both configuration-related issues and missing setup—similar to the second fuzz blocker in the motivating example. Specifically, line 3 captures a transformation flag defined via macros that specify the image transformation to apply (sub-type 2.2.3). Line 6 involves a set of macros that define the target image format the library should use when reading the input, thereby configuring the transformation from the input image to the expected format (sub-type 2.1).

We report the number of fuzz blockers matching the patterns of the sub-types supported by Query-3 in the 6\textsuperscript{th} column, and the number of fuzz blockers resolved in the 7\textsuperscript{th} column. In total, 36 fuzz blockers follow these patterns, and our query successfully resolves 29 of them (80\%). In other words, the query effectively suggests the missing function calls (sub-type 2.2) and missing valid argument options (sub-type 2.1).

We updated the fuzz drivers by inserting the \insightql{}'s suggested missing function calls or modifying function arguments in the existing fuzz drivers. We then recompiled the targets and executed the fuzzers for 24 hours, measuring code coverage improvements against the baseline (i.e., the original fuzz drivers). The results are shown in the 8\textsuperscript{th} column, with notable improvements observed in the \texttt{FreeType2} and \texttt{LibPNG} libraries. As a by-product of this process, we uncovered two bugs in \texttt{FreeType2}: one was a duplicate (previously reported), while the other was a previously unknown issue.\footnote{https://gitlab.freedesktop.org/freetype/freetype/-/issues/1336}

\begin{table}[ht]
  \centering
  \caption{Comparison of Line and Function Coverage in LibPNG: Manual fuzz blocker solving vs \insightql{}-assisted fuzz blocker solving}
  \label{tab:coverage_compare}

  \begin{tabular}{lccc}
    \toprule
    \textbf{Fuzz drivers} &
    \textbf{Line coverage} &
    \makecell{\textbf{Function cov.}\\\textbf{in \texttt{pngrtran.c}}} &
    \makecell{\textbf{Function cov.}\\\textbf{in \texttt{pngtran.c}}} \\ \midrule
    OSS\_Fuzz & 39.80\% & 52.08\% (25/48) & 19.05\% (4/21) \\
    Manual Effort            & 47.78\% \tablefootnote{We reproduced a lower coverage due to LibPNG code restructure. }& 60.04\% (29/48) & 19.05\% (4/21) \\
    \insightql{}       & 53.70\% & 75.00\% (36/48) & 52.38\% (11/21) \\ \bottomrule
  \end{tabular}
\end{table}

To further demonstrate the effectiveness of \insightql{} in addressing input-independent fuzz blockers, we compare our results with the manual analysis reported in~\cite{gao2023beyond}, which focused on unblocking fuzz blockers in LibPNG. Using \insightql{}, and in particular Query-3, we identified 17 unset flags or configurations that were blocking code paths—significantly more than the 6 identified through manual effort. This improvement led to a notable increase in code coverage for the LibPNG library, as shown in Table~\ref{tab:coverage_compare}.

\textbox{\textbf{Answer to RQ-2:} \insightql{} successfully analyzes 44.07\% and resolves 9.83\% of the studied fuzz blockers, resulting in code coverage improvements across six libraries. Notably, the improvements are particularly substantial in FreeType2 and LibPNG, with coverage increases of 10.17\% and 13.90\%, respectively. In the case of LibPNG, \insightql{} outperforms the manual effort reported in~\cite{gao2023beyond}, demonstrating its effectiveness in automating the identification and resolution of input-independent fuzz blockers.}

\subsubsection{Answer to RQ-3 regarding the usability of \insightql{}}
The objective was to measure and compare the time each participant spent on key tasks while analyzing fuzz blockers in the LibPNG library. The study focused on four key stages of an analysis using \insightql{}:

\begin{itemize}

\item \textbf{Setting up the \codeql{} database}: Participants set up the Visual Studio Code environment and created an \insightql{} database.

\item \textbf{Configuring the fuzzing environment}: Participants compiled a fuzz driver currently used in OSS-Fuzz.

\item \textbf{Using InsightQL to identify and classify fuzz blockers}: Participants used a coverage report generated by \fuzzintrospector{} and interacted with the codebase via the InsightQL extension in VS Code, analyzing fuzz blockers related to missing function calls using the query shown in Listing~\ref{lst:main-query-1}.

\item \textbf{Developing a new fuzz driver and conducting local testing}: Participants wrote a series of fuzz drivers incorporating the missing function calls, then performed local testing to compile the code and resolve any compilation errors or other issues.

\end{itemize}

\begin{figure}[ht]
  \centering
  \includegraphics[width=\columnwidth]{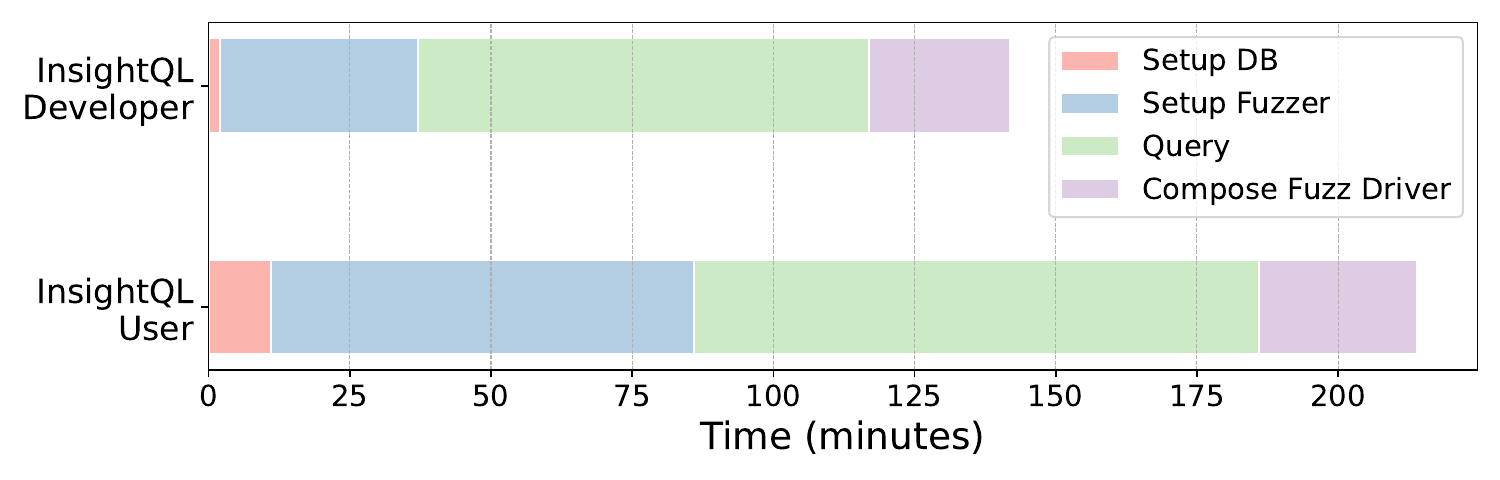}
  \caption{Comparison of time spent on each task between \insightql{} users and \insightql{} developers}
  \label{fig:user_study}
\end{figure}

As shown in Figure~\ref{fig:user_study}, compared to the developer of \insightql{} (the first author), the \insightql{} user (the forth author) spent approximately 100\% more time setting up the fuzzer but only 29\% more time interacting with the database and query interface.

The user reported two sub-tasks that took longer to complete. First, the user spent a longer time understanding how fuzzer is compiled under OSS-Fuzz's infrastructure. Second, the user spent an additional 20 minutes becoming familiar with the \insightql{} VS Code extension—specifically, interacting with panels, context menus, and interpreting query results.

The study also observed that both participants experienced some waiting time during query execution and encountered a brief learning curve when adapting to the workflow. While preliminary, these observations suggest that \insightql{} may offer good usability, potentially supported by its VS Code integration, generalized query templates, and intuitive right-click-and-query design. Overall, the results provide initial evidence that \insightql{} can assist developers of varying experience levels in introspecting fuzzing campaigns and addressing common analysis challenges.

\textbox{\textbf{Answer to RQ-3:} This preliminary user study suggests that \insightql{} is intuitive and offers a good level of usability, and can be used effectively by individuals with a basic understanding of C/C++ and fuzzing.}

\section{Threats to Validity} \label{sec:validity}

\textbf{Internal Validity.} Our answer to RQ-3 may be affected by the limited scope of the user study, which involved only two participants (both co-authors). While their feedback was valuable for initial validation, the generalizability of the findings is limited. A larger-scale study is planned, pending ethics approval. However, as both the approval process and participant recruitment require additional time, we intend to report the results of that study separately from this paper.

\textbf{External Validity.} The experiments were conducted on a selected set of real-world libraries and may not generalize to all software projects or domains. However, we chose widely-used and diverse targets to mitigate this risk.

\textbf{Construct Validity.} Our definition of ``fuzz blocker" and the criteria used to determine whether a blocker is resolved may not capture all real-world blocking behaviors. We followed prior work~\cite{gao2023beyond} to maintain consistency and used empirical metrics like code coverage to support our conclusions.
\section{Related work} \label{sec:related}
This section reviews prior research relevant to our work, spanning the representation of source code in queryable databases, approaches to human-assisted fuzzing, tools for analyzing fuzzing results, and efforts to leverage \codeql{} for fuzzing support. These areas collectively inform and motivate the design of \insightql{}.

\textbf{Code as a Database.} The representation of source code as data has traditionally relied on structures such as Abstract Syntax Trees (ASTs), Control Flow Graphs (CFGs), and Data Flow Graphs (DFGs). Yamaguchi et al. \cite{yamaguchi2014modeling,yamaguchi_automatic_2015} introduced the concept of combining these three graphs into a unified Code Property Graph (CPG) for static analysis, which became the foundation of the state-of-the-art static analysis tool Joern \cite{joern_io_Joern_The_Bug_2024}. Other notable tools that represent source code in a database format include Infer \cite{infer}, which uses a relational database to model control flow graphs, and the platform by Weiss et al. \cite{weiss2022languageindependentanalysisplatformsource}, which employs a graph database for CPG representation. CodeQL stands out for its extensive support across multiple programming languages and its rich query libraries. To the best of our knowledge, none of the existing tools integrate dynamic run-time information from fuzzing, which means they may have a limited view of the program under test.


\textbf{Human-assisted Fuzzing.} Human-assisted fuzzing refers to the incorporation of human expertise at any stage of the fuzzing process. Some notable human-assisted fuzzing techniques include manually crafting input to bypass fuzz blockers \cite{shoshitaishvili2017rise, brandt2024mind}; designing or modifying input grammars for generation-based fuzzing \cite{dutra2023formatfuzzer, pham2019smart, aschermann2019nautilus, wang2019superion}; annotating source code with hints to address difficult path constraints \cite{aschermann2020ijon}; and analyzing fuzz blockers to develop more effective test harnesses \cite{gao2023beyond, chanoss}. Recently, Fang et al. \cite{fang2024ddgf} introduced a dynamic directed greybox fuzzing approach to facilitate collaboration between developers and the fuzzer. However, unlike \insightql{}, their method requires humans to rank seed inputs in the corpus based on profiling results, without providing additional support. Moreover, the profiling is performed only on the seeds within the corpus, whereas \insightql{} collects a larger set of data from all executions.

\textbf{Human-assisting tooling for fuzzing result analysis.} There has been limited research in this critical area. Zhou et al. \cite{zhou2019visfuzz} introduced VisFuzz, a tool that visualizes and analyzes fuzzing progress at the call graph and control flow graph levels. Naik et al. \cite{naik2021sporq} developed SPORQ, an interactive code search tool that identifies similar design patterns when a user highlights program elements, and implemented a corresponding VS Code extension. \fuzzintrospector{} \cite{fuzzintrospector} is a recent open-source tool that visualizes fuzzing campaign coverage using function call graphs and control flow graphs, essentially encompassing the primary features of VisFuzz. Additionally, \fuzzintrospector{} assists in identifying fuzz blockers and estimating the potential code coverage improvement from unblocking them. \insightql{} leverages reports from \fuzzintrospector{} to help developers analyze these results more systematically and comprehensively.

\textbf{Using \codeql{} to Assist Fuzzing.} Since static analysis and dynamic analysis complement each other, researchers have explored using \codeql{} to support fuzzing. For instance, Liu and Videzzo \cite{liu_videzzo_2023} defined patterns that identify intra-message taints using CodeQL's taint tracking capabilities. Ebrahim et al. \cite{ebrahim2022fuzzingdriver} created regular expression queries with CodeQL to extract magic strings for fuzzing dictionaries. \insightql{} addresses an orthogonal problem: it assists fuzzing users in overcoming fuzz blockers in a more targeted manner.

\section{Conclusion} \label{sec:conclusion}

In this work, we highlight the critical need for human-assisting tools in fuzzing, particularly for analyzing and resolving fuzz blockers—one of the key obstacles to achieve higher code coverage. To address this gap, we introduce \insightql{}, the first integrated framework designed to assist humans in fuzzing analysis through a combination of static and dynamic insights. Built on Kimball's data warehouse design principles and powered by \codeql{}, \insightql{} features a unified code database and a flexible, parameterized query interface. Our evaluation across 14 widely-used libraries shows that \insightql{} can significantly aid in identifying and resolving fuzz blockers, leading to meaningful improvements in code coverage (up to 13.90\%). These results demonstrate the practical value of human-assisting tooling in enhancing fuzzing campaigns and ultimately improving software security. We hope \insightql{} serves as a foundation for future research in human-in-the-loop fuzzing and intelligent analysis support.

\section{Acknowledgement}

We thank Dongge Liu, Oliver Chang, and Cornelius Aschermann for their valuable feedback on this work. We also thank the anonymous reviewers for their insightful comments. This work was supported by the Australian Research Council’s Discovery Early Career Researcher Award (DECRA), project numbers DE230100473 and DE230100366, and also partly supported by the Institute of Information \& Communications Technology Planning \& Evaluation(IITP) grant funded by the Korea government(MSIT) (No.RS-2024-00440780, Development of Automated SBOM and VEX Verification Technologies for Securing Software Supply Chains).

\bibliographystyle{ieeetr}
\balance{}
\bibliography{references}

\end{document}